# Flow simulation in a 2D bubble column with the Euler-Lagrange and Euler-Euler method

Andreas Weber[a] and Hans-Jörg Bart[a]

[a]Chair of Separation Science and Technology, TU Kaiserslautern, PO box 3049,

andreas.weber@mv.uni-kl.de

**Abstract**

Bubbly flows, as present in bubble column reactors, can be simulated using a variety of simulation techniques. In order to gain high resolution CFD methods are used to simulate a pseudo 2D bubble column using EL and EE techniques. The forces on bubble dynamics are solved within open access software OpenFOAM with bubble interactions computed via Monte Carlo methods. The estimated bubble size distribution and the predicted hold-up are compared to experimental data and other simulative work using EE approach and show reasonable consensus for both. Benchmarks with state of the art EE simulations shows that the EL approach is advantageous if the bubble number stays at a certain level, as the EL approach scales linearly with the number of bubbles simulated. Therefore, different computational meshes have been used to also account for influence of the resolution quality. The EL approach indicated faster solution for all realistic cases, only deliberate decrease of coalescence rates could push CPU time to the limits. Critical bubble number - when EE becomes advantageous over the EL approach - was estimated to be 40.000 in this particular case.

## 1. Introduction

Bubble columns can be found in a various set of operations in chemical and biological applications. In order to reach a high efficiency, an optimal bubble size distribution (BSD) in any reactor type has to be achieved, which markedly depends on the local hydrodynamics (Akita and Yoshida, 1974). Computational fluid dynamics (CFD) simulation is one possibility of obtaining time and space resolved information in respect to complex bubbly flows. In general, they can be simulated in various ways, from detailed resolved single bubble behavior up to large-scale averaging methods with reactors with billions of bubbles inside. Special classes are based on particle population balances (PPB) relying on different break-up and coalescence kernels (B&C), that have been developed in the last few years to characterize the particle interactions (Liao and Lucas, 2009, 2010; Martínez-Bazán, 1998).

This work focuses on the Euler-Lagrange (EL) approach, giving a high level of detailed information on the bubble scale. In contrast to Euler-Euler (EE) simulations where the dispersed phase is considered as pseudo-continuous, every single bubble is resolved and can be tracked on its way through the domain. The EL approach makes use of stochastic modeling and improved collision algorithms to compete against other simulation techniques. All results are compared to solutions coming from state of the art EE simulations. Further, we investigate the computational cost of both (EL & EE) and identify critical parameters.

## 2. EL modeling

The EL approach for simulation of bubbly flows can be ranked between the direct numerical simulation (DNS) and the EE approach. While the EE model resembles the bubbles as a density distribution, in the EL model every single bubble is calculated as a point volume acting under Newtonian dynamics, while the movement of the surrounding fluid is solved via Navier-Stokes equations on an Eulerian grid. The internal coordinates of each bubble (size, concentration, etc.) are calculated using macroscopic models, which is the main difference to DNS simulations. However, the combination of this Eulerian grid and Lagrangian points yields a higher level of detail with a moderate increase in computational cost.

The original OpenFOAM solver *sprayFoam* provided the basis for the bubbly flow simulation shown in this work. *sprayFoam* uses an EL approach to simulate liquid or solid particles inside a continuous phase, e.g. a spray injected through a nozzle. The simulation of bubbles – gaseous particles – require a rework of this original solver to account for the additional physical effects.

### 2.1. Continuous phase hydrodynamics

Continuous phase and dispersed bubbles can be calculated predominantly independent from each other, which allows a solution in two steps. To achieve coupling of phases in the second, the exchange of momentum and energy from the bubbles is calculated and condensed into the Navier-Stokes source term $f$. With the continuous phase velocity $u_c$, the pressure $p$, the density $\rho$ and the viscosity $\mu$, the modified Navier-Stokes equation reads as:

$$\rho \left(\frac{\partial u_c}{\partial t} + (u_c \nabla) u_c\right) = -\nabla p + \mu \Delta u_c + f \qquad (1)$$

Turbulence is computed with Reynolds-averaged Navier-Stokes (RANS) modeling, which requires significantly less computational effort than solving for a Large Eddy Simulation (LES) approach. In the RANS model, only the averaged turbulence properties, like turbulent energy and dissipation, are calculated, while the LES approach will give a more detailed shape and movement of the actual turbulent eddies in the domain. A single phase k – epsilon turbulence model was chosen to calculate for the turbulence in the continuous phase and in addition to the original functionality, bubble induced turbulence (BIT) models have been added. The drag force, $F_D$, of a rising bubble with velocity $u_b$ induces turbulent energy and dissipation into the surrounding phase. This can be calculated using a source term for the turbulent $S_k$ and dissipative energy $S_\varepsilon$ equations:

$$S_k = \sum F_D \, |u_c - u_b| \tag{2}$$

$$S_\varepsilon = \frac{C_\varepsilon S_k}{\tau} \tag{3}$$

The open literature reveals different constants, $C_\varepsilon$, and turbulent time scales, $\tau$, according to specific physical effects in the bubbly flow (Pfleger and Becker, 2001; Rzehak and Krepper, 2013; Yao and Morel, 2004). In the presented EL simulation, the model of Rzehak and Krepper (2013) was used.

$$\tau = \frac{d}{\sqrt{k}}; \quad C_\varepsilon = 1.0 \tag{4}$$

**2.2 Bubble Dynamics**

Since the continuous phase solution has been vastly modified, the dispersed phase hydrodynamics yield a much larger rework of the original OpenFOAM solver. Nevertheless, the basic model of Lagrangian particle tracking persists. The bubbles are assumed to be points of mass at the position $x_b$. Local mass can change due to break-up and coalescence events. The new position of any specific bubble is calculated discretely with the velocity $u_b$.

$$\Delta x_b = u_b \Delta t \tag{5}$$

Forces acting on the bubble will change its velocity. Here $m_b$ resembles the bubble's mass and $\Sigma F$ stands for the sum of all acting forces on a single bubble.

$$m_b \Delta u_b = \Delta t \, \Sigma F \tag{6}$$

The sum of forces $\Sigma F$ consists of the buoyancy and weight force $F_B$, the drag force $F_D$, the lift force $F_L$, the virtual mass force $F_{VM}$, the wall lubrication force $F_W$ and the dispersion force $F_{TD}$. Here the subscripts $b$ and $c$ stand for the bubble and the continuous phase accordingly, the subscript $rel$ identifies the relative differences between them. Furthermore, g denotes the gravitational acceleration, $\rho$ stands for densities, u for velocity, $d_b$ for the bubble's diameter, $k$ for the turbulent kinetic energy and $\alpha$ for the phase fraction. Adjustable model parameters are denoted with $C_i$.

$$\Sigma F = F_B + F_D + F_L + F_{VM} + F_W + F_{TD} \tag{7}$$

These forces are in particular:

$$F_B = m_b g \left(1 - \frac{\rho_c}{\rho_b}\right) \tag{8}$$

$$F_L = m_b \frac{\rho_c}{\rho_b} C_L u_{rel} \times \nabla \times u_c \tag{9}$$

$$F_D = \frac{3}{4} C_D \frac{m_b \rho_c}{d_b \rho_b} |u_{rel}| u_{rel} \tag{10}$$

$$F_{VM} = -C_{VM} \rho_c V_b \left(\frac{D_b u_b}{Dt} - \frac{D_c u_c}{Dt}\right) \tag{11}$$

$$F_W = -C_W V_b \alpha_b \rho_c u_{rel}^2 n_{wall} \tag{12}$$

$$F_{TD} = -C_{TD} \rho_c k_c \nabla \alpha_b \tag{13}$$

In (11), $V_b$ stands for the bubble's volume and $Di/Dt$ denotes the material derivative, meaning that the derivative is made while following the bubble. In (12) the wall force parameter $C_W$ is dependent on the distance and relative velocity of the bubble to the next wall, $n_{wall}$ stands for the normal vector on the wall. In prior validation simulations (Weber and Bart, 2016) several models for the drag and lift force coefficients $C_D, C_L$ have been tested. Based on this, the models of Tomiyama (2004) were chosen. The virtual mass coefficient is set to $C_{VM} = 0.5$ according to Delnoij et al. (1997), the coefficient for the dispersion force is set to $C_{TD} = 0.1$ (Lahey et al., 1993).

Besides the movement and position information, each bubble is carrying an extra set of inner coordinates, in our case age, diameter, orientation, temperature and concentration of a solute. Additionally, physical properties like density, heat capacity or surface tension are calculated, which change with the surrounding pressure, temperature and solute concentration. However, in this initial simulations chemical reactions and mass transfer are neglected (no solute), which is a topic for further studies.

### 2.2.1. Stochastic modeling

Many aspects of bubble behavior are simulated using stochastic Monte Carlo (MC) models. On the one hand, it resembles the actual random character of bubble movement or other unpredictable interactions, and on the other hand it yields an immense benefit in computational effort. The procedure is similar in all situations; a calculated probability is compared to a random number which will trigger the event or neglect it. Through the law of large numbers, this algorithm will converge to the expectation value step by step. This seems a bit oversimplified but it represents a reliable method to calculate very complex systems, without the need of high dimensional differential equations.

#### 2.2.1.1. Bubble collision

The first bubble interaction, which is described through a stochastic model, is the collision between bubbles. Here the collision probability by O'Rourke (1981) is used. The actual position of bubbles is not considered, but rather the relative velocity and sizes of the bubbles are taken into account:

$$P_{coll} = \frac{\pi}{4}(d_{b1} + d_{b2})^2 |u_{b1} - u_{b2}| \frac{\Delta t}{V_{cell}} \qquad (14)$$

Note, that the volume of the computational cell $V_{cell}$ is also a parameter in this equation. This implies, that only bubbles belonging to the same cell can actually collide. Also, this has to be evaluated for every possible pair of bubbles in the interesting volume. For each pair, a random number $\xi \sim U(0,1)$ will be drawn. The constraint for collision will therefore be:

$$P_{coll} > \xi \qquad (15)$$

Compared to deterministic models, where the precise trajectory of every bubble has to be predicted, this stochastic approach is much faster. Because the collision step is one of the largest computational efforts in the code, it is crucial to use an efficient algorithm. One promising approach shown by Sigurgeirsson et al. (2001) was customized to work also with the stochastic collision algorithm gaining a major speed-up in comparison to the original code. The algorithm first creates a list of bubbles per each computational cell. In the second step, all combinations of bubble pairs for each cell are evaluated. Thereby, only bubbles in relative proximity will be observed. Moreover, the optimal cell size for this algorithm has been identified as the maximum bubble diameter. Smaller cells would presume a bubble, that has more volume than the enclosing cell, leading to numerical instabilities. However, a larger cell leads to more possible bubble pair combinations necessary to be calculated and slows down the process.

#### 2.2.1.2. Bubble coalescence

After a successful collision, the next step is to calculate the coalescence probability. Previous validation simulations (Weber and Bart, 2016) have shown that the model of Coulaloglou and Tavlarides (1977) is an appropriate choice. Basis of the calculus is the combination of bubble contact time $t_{contact}$ and film drainage time $t_{drainage}$. The drainage time can be seen as the time it takes for a thin liquid film between two bubbles to flow out. Only when bubbles are in long enough contact their surfaces can merge and coalescence happens. The time for contact is evaluated based on turbulence energy, $\varepsilon$. A normal distributed turbulent energy is assumed, which gives the average time that the bubble gets hit by a turbulent eddy, thus pushing them away from each other again. Analog to the collision, a new random number will be used for the constraint of coalescence:

$$t_{drainage} = C_{Cou,C} \frac{\mu_C \rho_C d_{eq}^4}{16\sigma^2} \left(\frac{1}{h_f^2} - \frac{1}{h_i^2}\right) \varepsilon^{2/3} (d_1 + d_2)^{2/3} \tag{16}$$

$$t_{contact} = \varepsilon^{-1/3} (d_1 + d_2)^{2/3} \tag{17}$$

$$P_{coa} = exp\left(-\frac{t_{drainage}}{t_{contact}}\right) \tag{18}$$

Here $\sigma$ represents the surface tension, $h_f$ and $h_i$ are critical and initial film thicknesses, respectively, and $d_{eq}$ is the equivalent diameter defined as:

$$d_{eq} = 2\left(\frac{1}{d_1} + \frac{1}{d_2}\right)^{-1}$$

If no coalescence happens, the bubbles will bounce of each other and continue their movement. Successful coalescence will lead to the deletion of one bubble, while its mass is added to the other one.

**2.2.1.3. Bubble break-up**

When a turbulent eddy with enough energy hits a bubble, it is deformed and eventually splits into smaller bubbles. The smaller a bubble is the more energy is necessary to deform it. Thus, the break-up frequency of small bubbles is low. This has also been modeled by Coulaloglou and Tavlarides (1977) into a break-up frequency, $\omega_b$, and probability, $P_{break}$. Again, a random number will be drawn and compared in each time step for each bubble.

$$\omega_b = C_{Cou,B1} \left(\frac{\varepsilon}{d_b^2}\right)^{1/3} exp\left[-\frac{C_{Cou,B2}\sigma}{\rho_b \varepsilon^{2/3} d_b^{5/3}}\right] \tag{19}$$

$$P_{break} = \omega_b \Delta t \tag{20}$$

When a break-up event has been evaluated as positive, a new bubble has to be created. This new daughter bubble has the same inner dimensions like its mother bubble. Only volume, diameter and mass will be changed on a basis of the volume ratio $f$ of the two bubbles. This is a controversial point in break-up modeling, since many researchers claim that $f$ should follow various different distributions. The model assumption of Coulaloglou and Tavlarides (1977) is a normal distribution of $f$, where the ratio $f = 0.5$ has the greatest probability. To sustain numerical stability, the creation of bubbles with an extremely small diameter is excluded by restricting $0.01 < f < 0.99$.

**2.2.1.4. Turbulent bubble movement**

In order to achieve chaotic bubble movement trough turbulent eddies, the bubble velocity is overlaid with a Brownian motion. The magnitude of this turbulent velocity is capped by the current turbulent energy. The direction is chosen uniformly random.

$$|u_{turb}| = \mathcal{N}\left(0; \sqrt{k\frac{2}{3}}\right) \tag{21}$$

This will lead to a trembling movement of bubbles positioned in a high turbulence area representing a diffusional movement. The higher the turbulence, the faster the bubbles will spread. Note that this additional velocity is not coupled to the forces and will not affect the creation of turbulent energy or the probability of collision.

## 3. EE modeling

The common approach to simulate bubbly flows is the usage of multiphase system solvers. Here, the dispersed phase is also modeled as a continuum and no explicit bubble positions will be evaluated anymore. Thereby, it is not possible to track single pathways of bubbles or to even assign precise diameters to bubbles in a certain volume. Anyhow, when incorporating a suitable model for coalescence and break-up, a EE simulation can still give valuable information about the mean diameter, surface area and phase hold-up of the gas phase inside a bubble column. The EE simulation is also capable of simulating a much higher dispersed phase hold-up like it would be possible in the EL approach (Hlawitschka et al., 2016).

There are many subcategories of EE solvers, while this work will concentrate on the comparison of EL simulations to EE simulations, which are combined with different moment based PPB models. The *method of moments* (MoM) (Hulburt and Katz, 1964) describes the dispersed phase based on the characteristic moments of the BSD: number, diameter, surface area and volume. In principle, the moments movement is solved like an incompressible isothermal fluid, also respecting the conservation of mass and impulse. Further improvements of this basic method have led to the *quadrature method of moments* (QMoM) (McGraw, 1997) and *direct quadrature method of moments* (DQMoM) (Marchisio and Fox, 2005), where closure problems have been overcome. Those two models are solved with a multiple equation system. In contrast to that, the *one primary one secondary particle method* (OPOSPM) (Drumm et al., 2010) and *interfacial area transport equation* (IATE) (Ishii et al., 2005) are based on only one equation to be solved, which improves the CPU speed even further.

For the comparison, the simulations of Hlawitschka et al. (2016) and Buffo et al. (2013) will be used.

### 3.1. Bubble Dynamics

In contrast to EL simulations, no explicit bubble position or velocity is calculated in the EE approach. Instead of that, the dispersed phase moments are used and their velocity is supposed to be equal in a computational cell. Transport of the moments is similar to a scalar transport, no interface tracking/sharpening steps are used. Nevertheless, the bubble forces are included by using volumetric equivalents of the explicit equations. With the mean bubble diameter $d_{3x}$ - calculated via the tracked moments - the phase hold-up $\alpha_c$ & $\alpha_b$ and the mean relative bubble velocity $u_{rel}$ the resulting forces can be calculated for each computational cell. For the drag force, the following equation is used.

$$F_D = \frac{3}{4} C_D \rho_c \alpha_c \frac{\alpha_b}{d_{3x}} |u_{rel}| u_{rel} \tag{22}$$

Same procedure is done for the buoyancy force. Other forces were neglected in the given simulations by Hlawitschka et al. (2016) and Buffo et al. (2013). For the drag coefficient $C_D$, Hlawitschka et al. (2016) used the model of Tomiyama (2004), Buffo et al. (2013) used the model of Haberman and Morton (1956). Bubble induced turbulence and turbulent dispersion were neglected in the EE simulations, the standard mixture k-epsilon RANS model was used. The coupling of phases was realized with a source term similar to eq. (1).

### 3.2. Bubble interaction

In general, the interaction of bubbles can be calculated using the same model equations like in the EL simulation. The difference is a conversion from explicit Monte Carlo simulation of a single or a pair of bubbles to a discrete application on the moments. The calculated event probability is no longer compared to random numbers but will be converted to an event frequency instead. This implies that it is possible to get fractional values (e.g. 0.5 bubbles per computational cell), which is impossible in the EL approach. While bubble volume stays constant the combined loss and gain in bubble number, size and surface area is recalculated during the bubble interaction step. The source term for the population balance can thereby be written as:

$$S(L, x, t) = B^C(L, x, t) - D^C(L, x, t) + B^B(L, x, t) - D^B(L, x, t) \qquad (23)$$

Where B stands for birth and D for death of bubbles, exponents C and B stand for coalescence and break-up respectively. While the coalescence model in Hlawitschka et al. (2016) is adopted from Coulaloglou and Tavlarides (1977), the break-up model originates from Laakkonen et al. (2006) which uses a similar approach based on the turbulent energy dissipation. Break-up of bubbles is supposed to be always binary and resulting daughter bubbles are supposed to be equal in size. Models for break-up and coalescence used by Buffo et al. (2013) are both adopted from Laakkonen et al. (2006). Here, the daughter bubbles created by break-up are supposed to follow a beta distributed volume ratio $f$.

While in the EL approach every single bubble diameter is inserted into the MC simulation, the EE approaches use representative diameters for all bubbles in the considered computational cell. This can either be the Sauter diameter $d_{32}$, the mean diameter $d_{30}$ or nodes of the quadrature points when using multi equation MoM models. This simplification enables EE simulation to be less dependent on the bubble number, making it a feasible calculation even with high gas hold-up.

**4. Simulation case**

There exist experimental data (Cachaza et al., 2009; Díaz et al., 2008b; Díaz et al., 2008a) and simulations (Hlawitschka et al., (2016), Buffo et al.,(2013)) for a set-up described in Díaz et al. (2008b) consisting of a rectangular bubble column depicted in Fig. 1:

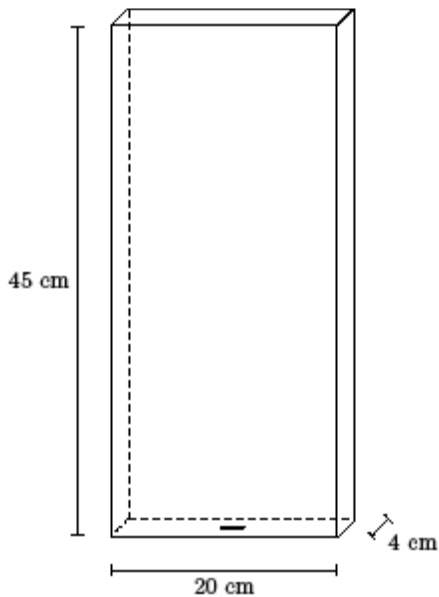

Fig. 1: Simulation case set-up

In the original experiment, the column is filled with tap water at room temperature while air is injected trough a sparger plate at the bottom. The sparger consists of eight holes with 1 mm diameter. They are arranged in two rows with a distance of 6 mm between holes.

In order to compare the two types (EE vs. EL) of simulation, the original EE mesh (Hlawitschka et al., 2016) has been used for the EL approach in a first simulation. To further improve simulation time and solution quality, an optimized mesh is used. The EE mesh consists of 24640 hexahedral cells with a length between 4 and 6.7 mm. Cells around the gas inlet zone are finer to account for a better solution. Since the EL simulation does not compute the free surface, the mesh is shortened to the filling height. The mesh refinement in the inlet area was turned off to achieve a more isotropic mesh. The resulting fine EL mesh consists of 11745 cells with a length between 6 and 7.5 mm.

The optimized coarse EL simulation case resembles the experimental volume within 20x4x45 computational cells, which gives uniformly sized cells of 1x1x1 cm. Note, that this coarse grid is inevitable in order to maintain a suitable ratio of maximum bubble size ($\approx$ 1 cm) and the grid cell length. This ensures that bubbles are always smaller in volume than the surrounding computational cell and that the collision of bubbles within a cell performs well. Other simulative work was built on a comparable coarse mesh, while Díaz et al. (2008b) even detected a numerical deviation when using a finer mesh. A grid convergence analysis was therefore disregarded for our EL simulation. Unlike EE Simulations, there is no need to define an inlet patch. The bubbles are injected on predefined points, depicting the eight sparger holes at the bottom.

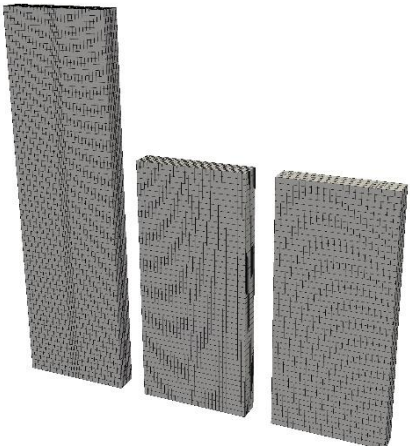

Fig. 2: Different meshes in comparison: left – original EE mesh; middle – simplified EE mesh; right – optimized coarse EL mesh

In experiment and simulation different gas flow rates have been tested. Corresponding velocities and bubble injection sizes have been calculated with equations from Geary and Rice (1991) and can be found in Table 1. Additionally, the *max+* case was simulated for the most promising models, although no experimental data exists for this case.

Table 1: Bubble sizes and number of bubbles to be injected for the different gas velocities.

| case name | min | med | max | max+ |
|---|---|---|---|---|
| superficial gas velocity | 2.4 m/s | 11.9 m/s | 21.3 m/s | 35.5 m/s |
| bubble inlet diameter | 6.3 mm | 8.3 mm | 10 mm | 10 mm |
| bubbles per second | 147 | 318 | 325 | 540 |

## 5. Results and discussion

Experimental and simulation results (s. Table 3) for Sauter diameter and hold-up are taken from Hlawitschka et al. (2016), Buffo et al. (2013) and (Díaz et al., 2008b). Note that in all cases, the hold-up is slightly under predicted in the EL in comparison to the EE simulations. The predicted Sauter diameter fits to the experimental values, except for the 'max' case, where it is markedly under predicted. The overall results of the EL simulation can be considered accurate enough in comparison with the experimental data and the other simulations. Values for the *max+* case are not listed, because no experimental results exist.

Table 3: Resulting sauter diameter, hold-up and plume oscillation time in comparison. [1]$d_{30}$

| case | Model / Ref. | Experiment / Diaz | DQMOM / Marchisio | OPOSPM / Hlawitschka | EL / own sim. |
|---|---|---|---|---|---|
| **min** | $d_{32}$ | 6.83 mm | 6.22 mm | 5.1 mm[1] | 6.17 mm |
|  | hold-up | 0.62 % | 0.62 % | 1.28 % | 0.52 % |
| **med** | $d_{32}$ | 6.5 mm | 6.69 mm | 5.9 mm[1] | 6.93 mm |
|  | hold-up | 2.63 % | 2.27 % | 2.86 % | 1.8 % |
| **max** | $d_{32}$ | 7.73 mm | 8.21 mm | 6.8 mm[1] | 5.52 mm |
|  | hold-up | 4.1 % | 4.06 % | 3.92 % | 4.2 % |

All simulations have been computed on the same machine (Core i7 4790, 16 GiB RAM, Ubuntu 14.04) to ensure equal conditions. Also, the simulations were done in single core mode, due to issues with some of the solvers in parallel mode. For all models, a total of 120 s experimental time has been simulated. Table 2 shows the resulting CPU time (in hours) in direct comparison. As expected, the more complex two equation models QMOM and DQMOM need significantly more time, while the one equation models IATE and OPOSPM yield a faster solution. The EL approach could easily beat the two equation EE models, while the one equation models were faster at lower gas throughput. When switching to the coarse mesh, the EL simulation could achieve solution in an even shorter duration, beating the simulation time of all EE simulations.

Table 2: Direct comparison of CPU time [h]

| Simulation type | EE | | | | EL | |
|---|---|---|---|---|---|---|
| model | QMOM | DQMOM | IATE | OPOSPM | Fine | coarse |
| **case** min | 10.99 | 4.95 | 3.03 | 3.27 | 4.22 | 0.47 |
| med | 33.10 | 33.82 | 14.72 | 15.52 | 7.82 | 2.78 |
| max | 43.91 | 33.57 | 19.79 | 14.56 | 11.26 | 3.33 |
| max+ | - | - | 20.3 | 14.09 | - | 5.18 |

Note the strong increase in computational time for the EE simulations from the *min* to the *med* case while the increase from *med* to *max* is not that significant. Simulations for the *max+* case showed no further significant increase in the computational time for higher gas velocities. Still, the EL simulation could achieve a faster solution for all listed cases and it does not seem useful to compare the models based on superficial velocity vs. CPU time.

A better approach (at least for the EL simulation) is the comparison of average bubble number and computational cost. I turns out that the computational time scales almost linearly to the average number of bubbles in the domain for the original coarse EL mesh (s. Fig. 3). Additional simulations were performed to prove this behavior even for a larger mesh. Therefore, the mesh dimensions have been doubled in every direction while resolution stayed the same, which gives a mesh with eight times more cells (*mesh x 8* in fig. 3). The superficial gas velocity has been further increased from 21.3 mm/s to 42.6 mm/s and 85.2 mm/s for this larger mesh.

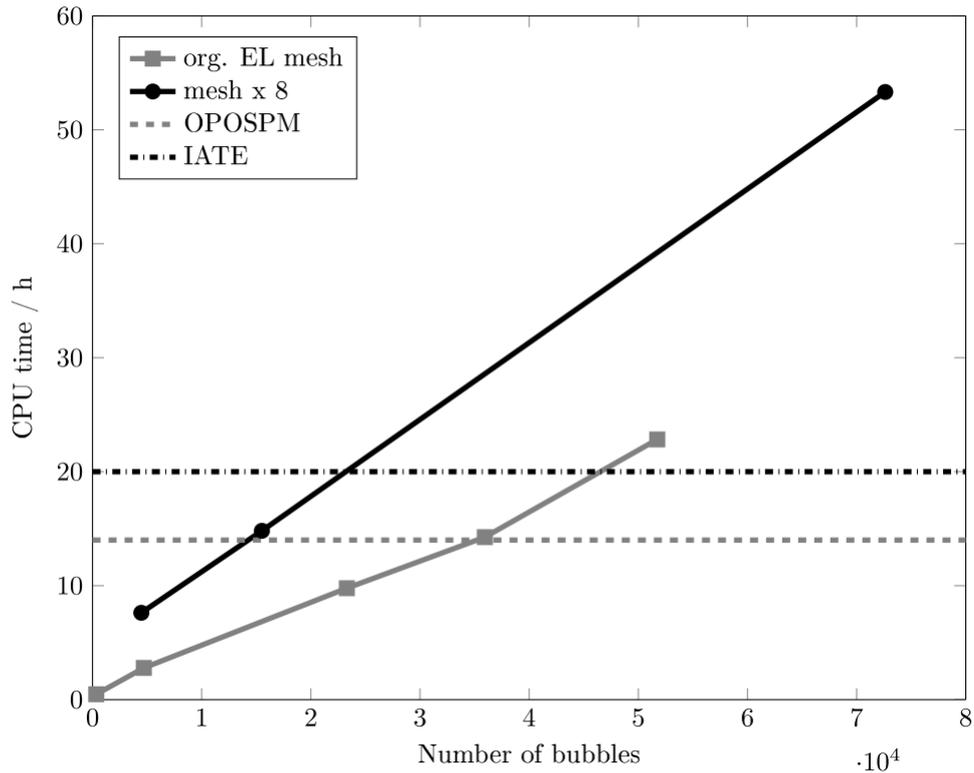

Fig. 3: Simulation CPU time for different numbers of bubbles in the domain

To find the critical number of bubbles, where the EE simulation finally becomes faster than the EL simulation, the coalescence efficiency was successively lowered. This leads to smaller bubbles and thus a higher number of bubbles. With an average bubble number of 36.000, the achieved computational time for the EL simulation reached a value of about 14 h (s. Fig. 4, grey dashed line), which is the time that the OPOSPM solver needed. A further increase to a bubble number of about 50.000 led to a simulation time of 23 h, which is even slower than the IATE solver. Thus, at least for this case geometry, the critical bubble number can be pinned to 40.000.

As mentioned above, a major speed up for the EL approach was achieved by using an efficient collision algorithm. According to literature (Sigurgeirsson et al., 2001), the optimal collision algorithm would need $n\ log(n)$ steps to calculate for collisions of $n$ particles. Since this can never be reached, the quality of an algorithm has to be quantified in terms of exponential notation $n^a$, where a smaller value for $a$ qualifies a better algorithm. Fig.3 compares the computational time of the original and improved algorithm that has been used.

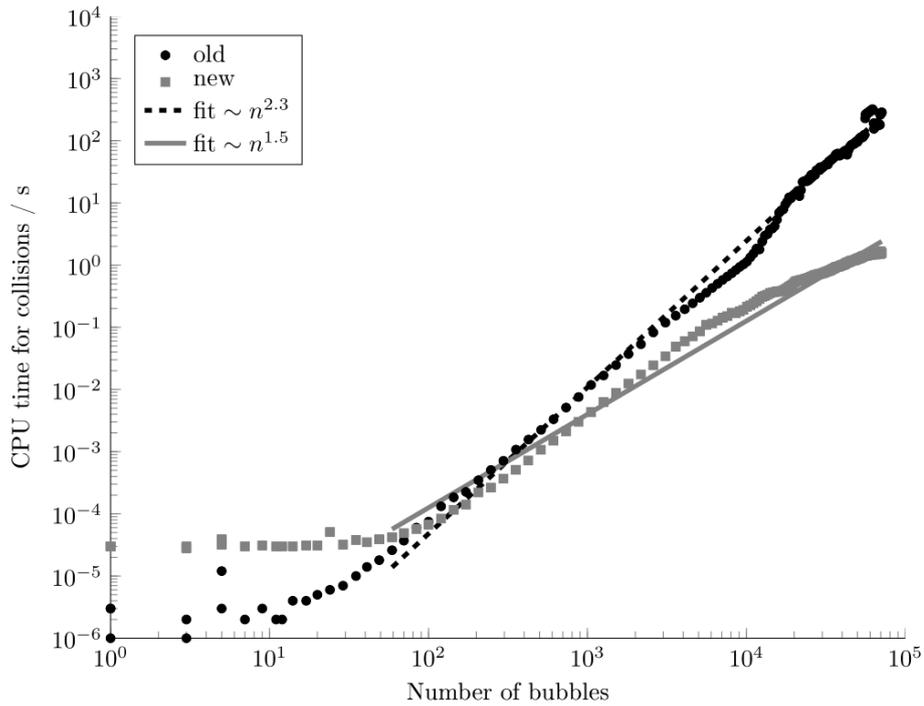

Fig.4: CPU time for colliding bubbles

Lowering the complexity from $n^{2.3}$ to $n^{1.5}$ is a large step: doubling the bubble number would have meant a five times higher CPU time but will now take only about 2.8 times higher effort with the new algorithm. The new one shows a slower solution for bubble numbers under 100 due to memory preallocation but its order for higher numbers shows a significantly faster calculation. Taking into account that in a bubble column are usually more than 100 bubbles, it is clear that the new approach is advantageous.

## 6. Conclusions

The EE and EL simulations are capable of solving the flow inside a bubble column, showing similar results in comparison with the experimental data. The computational time needed differs a lot from approximately one hour up to 40 hours, where a higher gas throughput corresponds to higher computational time. This dependency is existent even for the EE simulations but much more observable in the EL simulation. The EL algorithm is almost linear dependent on the number of bubbles and tends to be very slow, when the number increases to higher ranges ( > 40.000 bubbles). Here it is clearly advantageous to use an EE approach for an efficient simulation in cases, when the gas hold-up is high ( > 5%) or when the bubble size is small ($d_{32}$ < 4 mm). Still, there are many applications with a medium bubble size and/or low gas hold-up. In this case, the EL simulation technique can easily outperform the EE simulations in CPU time and level of detail. This is due to the scaling of the EL simulation to the number of bubbles, which is almost linear, while the EE technique has already high computational cost even at a relatively low hold-up. Computational cells with no bubbles present have almost no impact on CPU time in the EL approach. In the EE approach, even those cells have to be computed concerning transport and interaction of moments.

Further advantage of the EL approach is the easy visualization of trajectories and positioning of single bubbles on their way through the reactor. The EE simulation results are treated in post processing to reproduce such a meaningful visual representation although no "real" path of single bubbles were calculated. When concerning chemical reactions, detailed information about its temperature and concentration progress can also be combined with particle residence time, path length and other explicit single bubble properties within the EL approach.

**Acknowledgements**

Funding by the Deutsche Forschungsgemeinschaft (DFG) within the RTG GrK 1932 "Stochastic Models for Innovations in the Engineering Sciences", project area P1, is gratefully acknowledged. We want to thank Mark W. Hlawitschka, TU Kaiserslautern, for supply of the EE solvers used in this work.

**Nomenclature**

Latin letters

| | |
|---|---|
| $d$ | (bubble) diameter |
| $d_{23}$ | sauter diameter |
| $f$ | volume ratio |
| $F$ | force |
| $g$ | gravity |
| $h$ | layer thickness |
| $k$ | turbulent energy |
| $m$ | mass |
| $n$ | number; normal vector |
| $P$ | probability |
| $u$ | velocity |
| $rel$ | relative |
| $t$ | time |
| $p$ | pressure |
| $V$ | volume |
| $x$ | position |

Greek letters

| | |
|---|---|
| $\alpha$ | phase fraction |
| $\xi$ | random number |
| $\varepsilon$ | turbulent energy dissipation |
| $\sigma$ | interface tension |
| $\omega$ | frequency |
| $\rho$ | density |
| $\mu$ | viscosity |
| $\tau$ | turbulent time scale |

Subscripts, superscripts, symbols

| | |
|---|---|
| b | bubble |
| B | buoyancy; break-up |
| C | continuous phase; coalescence |
| D | drag |
| eq | equivalent |
| L | lift |
| VM | virtual mass |
| W | wall |
| TD | turbulent dispersion |
| turb | turbulent |
| k | turbulent energy |
| rel | relative |
| $\varepsilon$ | turbulent energy dissipation |

$\frac{D_i}{Dt}$     material derivative

$\mathcal{N}$     normal distribution